\documentclass[preprint, 11pt, a4paper, times,trackchanges]{aastex631}
\usepackage[english]{babel}

\usepackage{amsmath} 

\usepackage{graphicx}
\usepackage{natbib}
\usepackage{gensymb}

\newcommand{\lya}{Lyman-$\alpha$~{}}
\newcommand{\kms}{km s$^{-1}$~{}}

\begin{document}
\title{Sensitivity of the helioglow to variation of the total ionization rate and solar \lya emission.}

\correspondingauthor{I. Kowalska-Leszczynska}
\email{ikowalska@cbk.waw.pl}

\author[0000-0002-6569-3800]{I. Kowalska-Leszczynska}
\affil{Space Research Centre PAS (CBK PAN),\\
Bartycka 18A, 00-716 Warsaw, Poland} 

\author[0000-0002-5204-9645]{M. A. Kubiak}
\affil{Space Research Centre PAS (CBK PAN),\\
Bartycka 18A, 00-716 Warsaw, Poland}

\author[0000-0003-3957-2359]{M. Bzowski}
\affil{Space Research Centre PAS (CBK PAN),\\
Bartycka 18A, 00-716 Warsaw, Poland}

\author[0000-0003-3484-2970]{M. Strumik}
\affil{Space Research Centre PAS (CBK PAN),\\
Bartycka 18A, 00-716 Warsaw, Poland}

\begin{abstract}
Direct observations of solar wind are mostly limited to the vicinity of the ecliptic plane.
Retrieving the latitudinal structure of solar wind indirectly based on observations of the backscatter glow of interstellar neutral hydrogen is complex and requires support from theoretical models.
The GLOWS instrument, to operate on the planned IMAP mission, will scan the helioglow along circumsolar rings with an angular distance of  $\sim 75\degree$.
Its objective is to retrieve the latitudinal structure of the ionization rate of interstellar hydrogen and with this, the structure of the solar wind.
In preparation for the future analysis, we studied the sensitivity of the light curves to temporal and latitudinal variation of the ionization rate of interstellar hydrogen and the solar \lya illumination.
Based on carefully planned numerical experiments, we analyze the time delay and relaxation time of the system for variations of the ionization rate and solar illumination in heliolatitude and with time.
We found that variations in the solar illumination are reflected in the helioglow without delay, but relaxation takes longer than the variation rise time. 
By contrast, variations in the ionization rate are anti-correlated with the helioglow brightness with a delay of several months.
We also found that the helioglow is not sensitive to variations in the ionization rate at the solar poles, so retrieving the ionization rate and solar wind at the poles requires approximation of the ionization rate profiles with appropriate parametric functions.
\end{abstract}

\section{Introduction}
\label{sec:intro}
\noindent
The state of the immediate surroundings of the heliosphere is determined by a balance between the matter blown out from the Sun in the form of solar wind and the incoming partially ionized, magnetized interstellar matter.
Charged particles flow past the heliopause, while neutral atoms enter inside freely and penetrate the inner heliosphere.
The unperturbed interstellar medium, in which the heliosphere is immersed, currently can only be studied through analysis of neutral atoms that reach the vicinity of the Sun.

Along the way inside the heliopause, interstellar gas is subjected to ionization losses.
Ionization of interstellar neutral (ISN) atoms occurs by three mechanisms: charge exchange due to collisions with solar wind ions, and ionization by absorption of EUV photons or collisions with solar wind electrons.
Since the flux of the ionizing particles (solar wind protons and EUV photons) drops with the square of solar distance, the ionization rate follows this dependence. 
The rate of ionization by electron impact slightly deviates from this rule, but adopting a view where the ionization rate drops with the solar distance squared is a very reasonable approximation.
Hence, it can be conveniently expressed by the ionization rate at 1 au, understanding that it needs to be appropriately scaled for an actual distance. 

Taking into account these three mechanisms allows to determine the total ionization rate, which modulates the density of interstellar matter within the heliosphere.
For ISN H, the most important ionization process is charge exchange. 
Since the charge exchange rate is a function of the speed and density of the solar wind protons, the latitudinal structure of the solar wind and its evolution with the solar cycle phase is one of key factors shaping the distribution of ISN H.
During the solar cycle, the magnitude of the total ionization rate of ISN H varies mostly within the polar regions of the heliosphere \citep{bzowski_etal:13a, sokol_etal:19a} coherently with the phase of the solar cycle, with an amplitude $\sim 1.5$. 
At equatorial latitudes, these variations are milder and do not feature a clear coherence with the phase of solar cycle.

Another important factor is radiation pressure. 
The radiation pressure force is created due to resonant absorption of solar \lya photons and  resulting momentum absorption by the target H atom.
It is directed away from the Sun and drops with square of solar distance.
With this, it is conveniently expressed by a factor of (over)compensation of solar gravity force acting on individual H atoms.
Since the solar spectral flux within the \lya line features a characteristic double-horn shape, the radiation pressure factor $\mu$ is a function of radial speed of an individual H atom due to the Doppler effect.
Since the solar \lya flux varies with time, this compensation factor $\mu$ also varies with time during the solar cycle, with the horn to line center ratio equal to $\sim 1.4 - 1.7$, depending on the strength of a given solar cycle. 
The magnitude of the $\mu$ factor varies between $\sim 0.8$ and $\sim 2$ during the solar cycle.
Within a few au from the Sun, its significance for the density distribution of ISN H is at least as large as that of ionization \citep{tarnopolski_bzowski:09, IKL:18b}.
                  
The solar wind structure can be studied in two ways.
One of them is direct in-situ measurement of the density and velocity of the solar wind plasma.
These measurements are typically limited to the ecliptic plane, where most of the plasma instruments have been located.
An exception was the Ulysses mission \citep{marsden_smith:97}, whose orbit allowed for capturing particles even above the solar poles. 
This method provides ground truth, but its limitation is the point-like character of the measurements and the resulting inability to provide an instantaneous global view. 

The other way is to use one of two currently available indirect methods, i.e., remote-sensing observations either of interplanetary scintillations \citep[IPS, e.g., ][]{jackson_etal:20a, tokumaru_etal:21a} of remote compact radio sources on solar wind electrons, or of the heliospheric backscatter glow \citep{bertaux_etal:95}. In both cases, analysis requires additional assumptions and modeling, but provides a global view of all heliolatitudes simultaneously.

In this article, we focus on the relation between the heliospheric backscatter glow (further on: the helioglow) and the structure of the solar wind.
The helioglow is created due to re-emission of the solar \lya photons absorbed by ISN H atoms.
Clearly, the helioglow creation process is tightly related to the radiation pressure effect, but the helioglow distribution in the sky is an indirect tracer of the density distribution of ISN H, and thus also depends on the ionization rate.
The solar wind structure was successfully retrieved based on analysis of the helioglow by several authors, e.g., \citet{bertaux_etal:99,bzowski_etal:03a, lallement_etal:10b, katushkina_etal:19a}, and \citet{koutroumpa_etal:19a}. 
Most of these analyzes were made based on data from the SWAN experiment \citep{bertaux_etal:95} on board the SOHO mission  \citep{domingo_etal:95a}.
The objective of the upcoming GLObal solar Wind Structure (GLOWS, \url{https://glows.cbk.waw.pl/}) experiment, which will be part of the Interstellar Mapping and Acceleration Probe mission \citep[IMAP,][]{mccomas_etal:18b}, is to study the evolution of the solar wind structure based on analysis of helioglow light curves obtained for Sun-centered scanning circles.

The effects of the ionization rate, radiation pressure, an illumination of ISN H reflected in the helioglow are challenging to separate observationally. 
This article is the first step in understanding how sensitive the GLOWS-like light curves are to changes in the total ionization rate and radiation pressure.
We perform several numerical experiments aimed at understanding the sensitivity of the helioglow to the ionization rate of ISN H globally, as well as to its variations in carefully selected latitudinal bands. 
We want to find out how sensitive the light curves are to variations of the ionization rates in polar latitudes compared with variations at equatorial and middle latitudes, and what are the time scales of helioglow reaction to these variations.
We also investigate the sensitivity of the light curves to variation in the solar illumination and the related time scales.

These studies will help to identify the strategy of retrieving the solar wind structure from future GLOWS observations.
The widely used approach to reconstruct the ionization rate and solar wind structure based on post-processed observations relies on the assumption that the latitudinal structure of the ionization rate can be divided into independent latitude bands.
The total ionization rate profiles were obtained from fits of ionization rate profiles composed of independently-varying rates in pre-defined latitudinal bins. 
This approach was used by, e.g., \citet{summanen_etal:93} and several authors afterward. 
An alternative approach \citep[e.g.,][]{lallement_etal:85a, bzowski:03} is to adopt a functional representation of the ionization rate profile, with a few adjustable parameters.
This latter approach implicitly assumes that the solar wind varies on a global scale, making a homogeneous physical system. 
This seems to be a reasonable approach given the predominant role of the solar magnetism.
Allowing for free uncorrelated variations in the solar wind outflow implicitly assumes that the behavior of solar wind in one latitudinal band is uncorrelated with that in other latitudes.
This would be similar to adopting uninformed priors in a Bayesian analysis. 
Our research helps to select an analysis strategy for GLOWS.

In Section \ref{sec:sim_lc}, we introduce the IMAP/GLOWS instrument along with the GLOWS signal simulator package that allows to generate simulated light curves for various profiles of the ionization rate and solar \lya histories.
In Section \ref{sec:ion_models}, we present results of a series of numerical experiments examining the influence of changes in the total ionization rate.
We analyze the impact of the ionization rate intensity in the case of spherically symmetric ionization (Section \ref{sec:ion_stat_lat}), as well as effects of increased ionization at selected latitudinal bands (Section \ref{sec:latiVaria}). 
Then, we study a transient increase of the total ionization rate caused by the spherically symmetric solar wind changes or 
by the solar wind change where latitudinal profile is similar to realistic one (Section \ref{sec:ion_sw_time}).
Finally, in Section \ref{sec:lya_time}, we investigate the sensitivity of the helioglow to transient variation of the \lya{} illumination and resulting radiation pressure, and related variation timescales while keeping the total ionization rate constant (Section \ref{sec:lyaFlash}). 
We also analyze effects of an anisotropy of the solar illumination, pointed out by several authors \citep[][]{pryor_etal:92, pryor_etal:96, auchere:05, strumik_etal:21b, strumik_etal:24a} but frequently neglected in existing analyses (Section \ref{sec:lyaFlashLati}).
We summarize and conclude our research in Section \ref{sec:summary}.

\section{Simulation model and GLOWS light curve}
\label{sec:sim_lc}
\noindent
IMAP will operate in the ecliptic plane close to the first Lagrange point (L1) between the Earth and the Sun.
During each rotation of the spacecraft, GLOWS will scan a strip of the sky with a radius of 75$\degree$ around a point located 4$\degree$ in ecliptic longitude behind the Sun. 
The spin axis will be adjusted daily.
The resulting light curves will be modulated by variations in the total ionization rate of ISN H and solar \lya illumination. 
Additional modulation of the light curve is expected due to the fact that the flow direction of ISN H is inclined at a certain angle to the ecliptic plane. 
In all simulations presented in this paper, we consider an observer located at $\sim 1$ au from the Sun, i.e., approximately at the Earth orbit.
An example simulated light curve that we expect from the GLOWS detector (after removing all extra-heliospheric sources) for a realistic ionization rate is presented in Figure \ref{fig:map_lc}.

In this paper, we use a highly simplified version of the WawHelioGlow model \citep{kubiak_etal:21a}.
Even though \citet{baranov_malama:93} suggested that ISN H is expected to feature two populations, the primary and the secondary, \citet{kubiak_etal:21a} compared predictions from a two-population and one-population models and demonstrated that helioglow simulations at 1 au can be performed assuming a single population of ISN H with the inflow parameters obtained as the average of those of the primary and secondary populations: ecliptic longitude $\lambda=252.5 \degree$, ecliptic latitude: $\phi=8.9 \degree$, flow velocity: $V=21.26$ km s$^{-1}$, temperature $T=12\,860$ K, and density $n=0.127$ cm$^{-3}$.
These flow parameters are consistent with those obtained from analysis of spectroscopic observations of the helioglow by \citet{lallement_etal:05a, lallement_etal:10a}.

WawHelioGlow needs to use models of the ionization rate and radiation pressure. 
For radiation pressure, we adopt a model developed by \citet{IKL:18a, IKL:20a}. This model takes the total irradiance in the \lya line a radial speed of an atom and calculates the corresponding $\mu$-factor. 
In the calculations discussed in Section \ref{sec:ion_models}, the magnitude of the solar \lya irradiance was adopted as constant in time and equal to $3.8 \times 10^{11}$ photons cm$^{-2}$ s$^{-1}$, which is a typical value for low solar activity. 
In Section \ref{sec:lya_time}, we adopt a time-dependent model presented in this section.

The ionization rate in the WawHelioGlow code can be adopted as a sum of the rates of photoionization, charge exchange, and electron impact reactions, with the rates of charge exchange calculated from adopted profiles of the solar wind density and velocity. 
For this study, however, we adopt a total ionization rate without breaking it down into individual components.
Details of the ionization rate models used in Section \ref{sec:ion_models} are discussed in individual subsections.

The simulations presented in this paper have been performed for vantage points close to three cardinal points, near upwind, downwind, and crosswind.
Since, however, actual observations are never taken precisely at these locations, we adopted vantage points identical with three actual positions of SOHO in space near these points.
This, together with the 4\degree{} offset of the scanning circle from the Sun's center, results in small but significant departures from the perfect upwind, downwind, and crosswind viewing geometries.

\begin{figure*}[ht!]
	\includegraphics[width=1\linewidth]{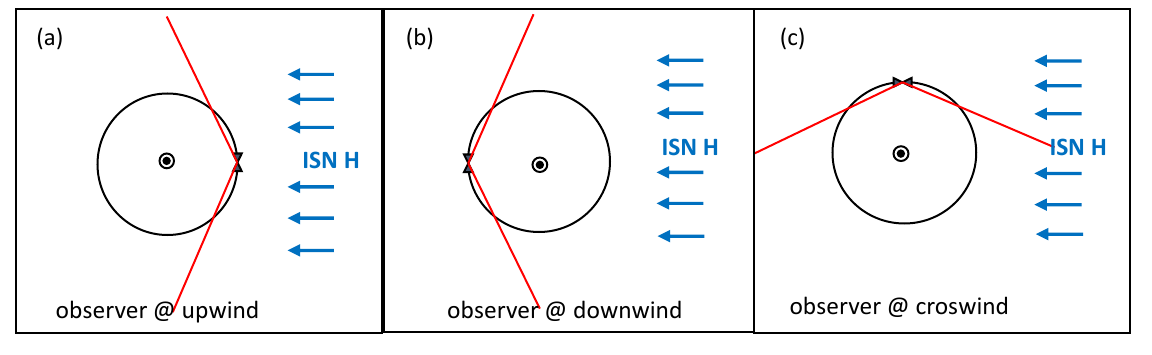}
	\caption{Viewing geometries in the ecliptic plane that were used in the simulations. Earth's orbit is shown as a black circle with the Sun marked in the center. Observer is represented by the black hourglass. Interstellar Medium flows from the right side of each panel (blue horizontal arrows). The red lines show two GLOWS's lines of sight located in the ecliptic plane. Spacecraft rotation causes the line of sight to trace a circle in a plane perpendicular to the ecliptic plane. The angular distance of the lines of sight from the point near the Sun, at which the satellite's rotation axis aims, is 75$\degree$.
	}
	\label{fig:obs_position}
\end{figure*}
 
The three vantage points adopted in the simulations are the following: upwind $\lambda=254.50 \degree$, $\phi=0.03 \degree$; downwind $\lambda=69.97 \degree$, $\phi=0.00 \degree$; and crosswind $\lambda=165.85 \degree$, $\phi=-0.04 \degree$. Positions of the observer with respect to the ISN H flow direction are shown in Figure \ref{fig:obs_position}.
These vantage points correspond to the actual SOHO positions nearest the nominal upwind, downwind, and crosswind positions in 2009.

\begin{figure*}[ht!]
	\centering
	\includegraphics[width=1\linewidth]{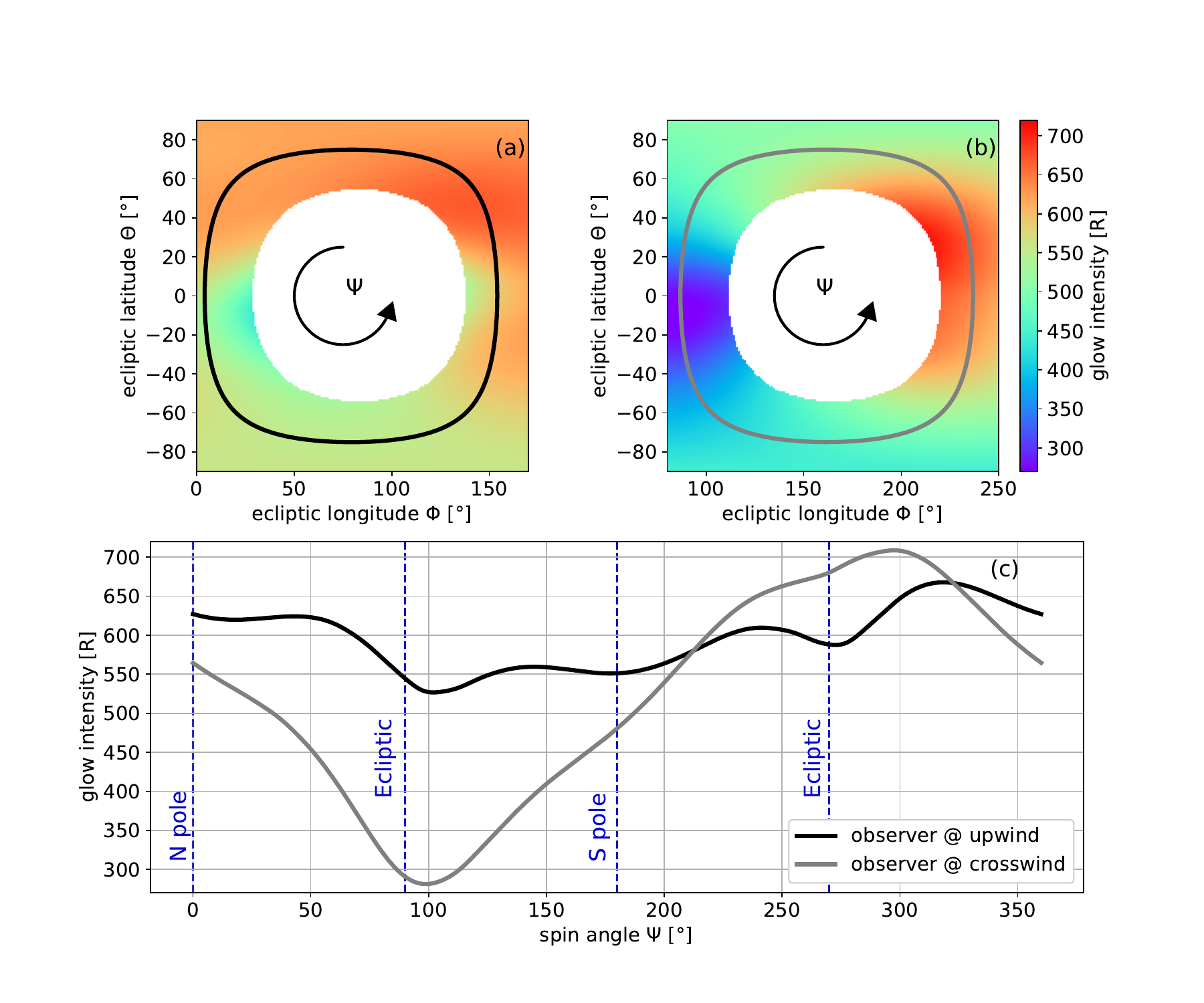}
	\caption{Simulations of the hydrogen helioglow seen by the observer at upwind (panel (a)) and crosswind (panel (b)). Spin angle $\Psi$ is counted from the north ecliptic pole counter-clockwise. Panel (c) shows light curves taken along the scanning circles shown on the helioglow maps above.
	The spin angle is counted counterclockwise off the northernmost point in the scanning circle.  The simulations were performed for June 14, 2009 (upwind) and September 8, 2009 (crosswind). The white area at the center of panel (a) and (b) corresponds to the exclusion region around the Sun.
	}
	\label{fig:map_lc}
\end{figure*}

Figure \ref{fig:map_lc} shows examples of our state-of-the-art simulations of sky maps of the hydrogen backscatter glow calculated using the WawHelioGlow model.
Panel (a) presents the view for an observer located in the upwind vantage point at 1 au and facing the Sun, thus having the inflow of ISN H from behind.
Panel (b) shows a side view of the ISN H flow, where the scanning circle passes through the upwind and downwind regions (at the right and left side in the panel, respectively), i.e., those featuring the highest contrast in the hydrogen density.
The colored background in these panels represents the intensity of the helioglow calculated on a regular grid of Sun-centered rings and bi-linearly interpolated. 
The GLOWS light curves shown in panel (c) are taken along the black and gray scanning circles shown in panels (a) and (b).
They have complicated shapes, which is a result of the non-trivial structure of the solar wind and the modulation specific to individual observer locations. 
Note that the contrast, i.e., the maximum to minimum ratio, is the largest for the crosswind light curve. This is because the scanning circle in this geometry passes through the regions with the maximum and minimum densities.

The light curves in Figure \ref{fig:map_lc}(c) were obtained by bi-linear interpolation of the background map of the helioglow. Interpolation was calculated for the geometric location of the scanning circles in the sky. 

Figure \ref{fig:map_lc} is intended to give a glimpse how realistic light curves look like for the hydrogen glow. 
More examples of light curves for different vantage points around the Sun are shown by \citet{bzowski_etal:23b} in their Figure 1.

For generation of the light curves presented in the next sections, a simpler procedure was used.
The simulations were performed directly for the directions forming the scanning circles, i.e., no interpolation method was necessary.

\section{Sensitivity of the light curves to various aspects of ionization rate profiles}
\label{sec:ion_models}
\noindent
Our objective in this section is to better understand the effect of ionization on helioglow light curves. To that end, we break down a complex problem into simple factors and investigate the impact of each of them on the simulated GLOWS light curves, which allows us to assess the chances of distinguishing them in real observations.
We focus on several aspects, which are described below.

\subsection{Sensitivity of light curves to absolute magnitude of the ionization rate, constant in time}
\label{sec:ion_stat_lat}

\noindent
The first question is how the overall helioglow intensity changes for various magnitudes of a spherically symmetric ionization rate and how the shapes of the light curves are affected.

\begin{figure*}[ht!]
\centering
\includegraphics[width=1.1\linewidth]{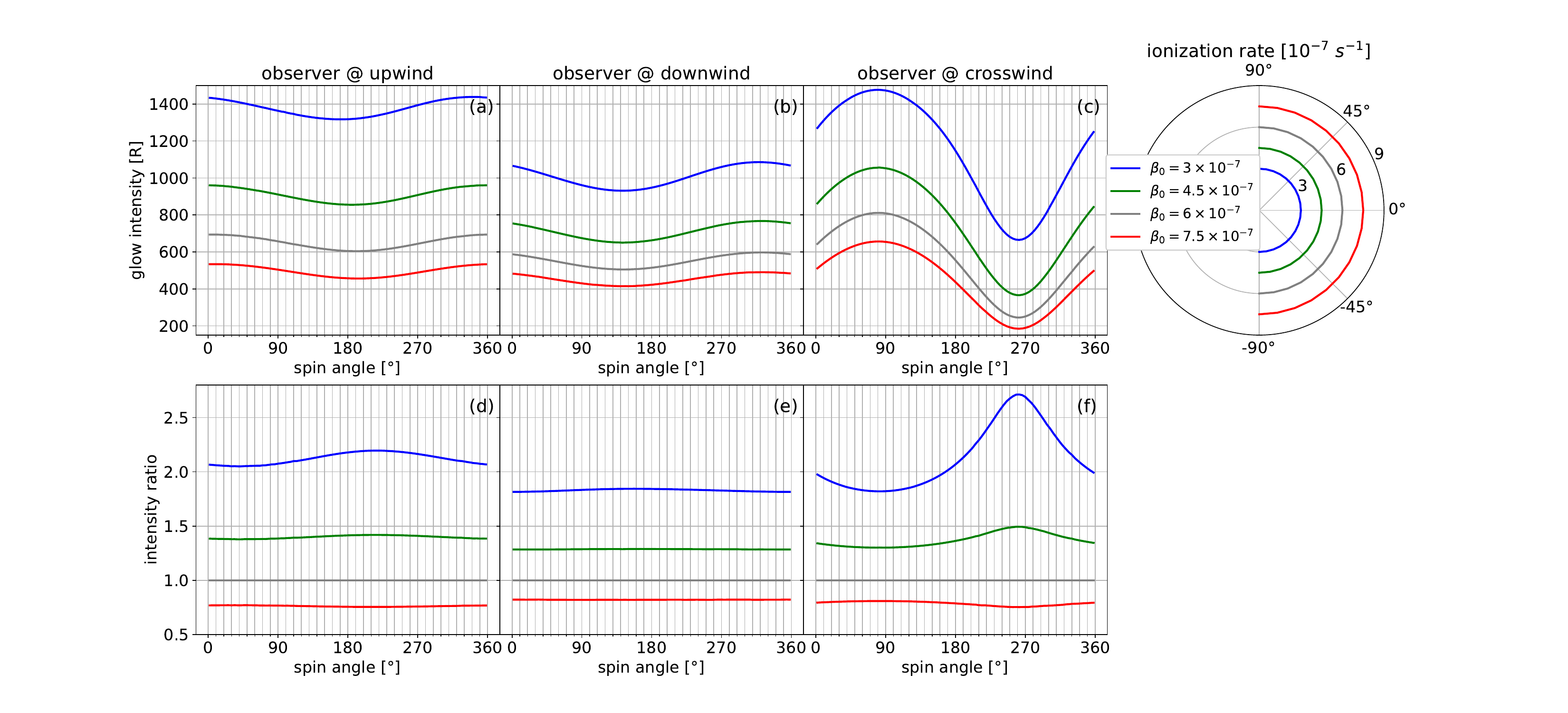}
\caption{Simulated GLOWS light curves (panels (a)---(c)) and their respective intensity ratio (panels (d)---(f)) compared to the case for $\beta_0=6 \times 10^{-7}$ s$^{-1}$, as seen from three vantage points: upwind, downwind, and crosswind for different global total ionization rates $\beta_0$ denoted by colors. Polar plot on the right shows the latitudinal profiles of the ionization rates for each light curve. In this case, all these profiles are flat, i.e., feature null dependence on heliolatitude.
}
\label{fig:ion_stat_beta0}
\end{figure*}

We perform four simulation runs for the upwind and crosswind locations, varying the parameter $\beta_0$ within a range characteristic for the global minimum and maximum of the ionization rate at the solar poles: $\beta_0=3 \times 10^{-7}$ s$^{-1}$, $4.5 \times 10^{-7}$ s$^{-1}$, $6 \times 10^{-7}$ s$^{-1}$, and $7.5 \times 10^{-7}$ s$^{-1}$ \citep[see ][]{sokol_etal:20a}. 
The baseline magnitude of the ionization rate is adopted as $\beta = 6 \times 10^{-7}$ s$^{-1}$. 
The polar plot in Figure \ref{fig:ion_stat_beta0} (the top right panel) shows the flat ionization rate profiles for this simple spherically symmetric model.
The upper three panels (a)---(c) in Figure \ref{fig:ion_stat_beta0} show the light curves, while the lower three panels (d)---(f) show the light curve ratios with respect to the baseline case, as seen by the observer in different vantage points: upwind, downwind, and crosswind. 
Even in this simple case, the light curves are not expected to be featureless.
This is because the detector is moving in a plane to which the flow of ISN H is inclined at a certain angle, while the flow, and thus the helioglow, are expected to feature axial symmetry around the flow direction.
Since the total ionization rate and solar illumination are spherically symmetric, the light curve modulation is a result of the anisotropy in the ISN H distribution due to streaming of the ISN H through the heliosphere. 

As expected, a decrease in the total ionization rate causes an increase of the helioglow intensity (panels (a)---(c) in Figure \ref{fig:ion_stat_beta0}), but not uniformly, which is clearly visible in panels (d)---(f). 
For the adopted range of the ionization rates, the change in the intensity is approximately logarithmic as a function of the ionization rate.
The parameters of this logarithmic dependence vary between the observer positions and between different spin angles. 
The largest spin-angle variations are for the crosswind vantage point.

For an observer placed near the upwind and downwind locations,
there is a small modulation of the light curve caused by the gas distribution (panels (a,b) in Figure \ref{fig:ion_stat_beta0}). 
With the flow coming from above the ecliptic plane, the minimum of the light curve is expected for spin angle 180\degr. 
However, the departures of the viewing geometry from the ideal case result in a small dependence of the position of the light curve minimum on the magnitude of the ionization rate, which is visible as the systematic shift of the position of the minimum to the right in panel (a) and to the left in panel (b).

The most complex shape of the light curve, as well as the largest maximum to minimum intensity ratio is seen by a crosswind observer. 
From this vantage point, the ISN H flow is seen sideways. 
Therefore, one part of the light curve, close to spin angle 90\degree, probes the denser gas in the upwind hemisphere, while the other, close to spin angle 270\degree, probes the downwind hemisphere, where the hydrogen densities are much lower (panels (c) and (f) in Figure \ref{fig:ion_stat_beta0}).
The maximum to minimum ratio for each light curve on panel (c) increases from $\sim 2.2$ to $\sim 3.5$ with the transition from $\beta = 3\times 10^{-7}$ s$^{-1}$ to $6 \times 10^{-7}$ s$^{-1}$.

\subsection{Sensitivity to ionization variations at isolated  latitudes}
\label{sec:latiVaria}
\noindent
In this section, we investigate the sensitivity of the light curves to variations in the ionization rate within narrow, isolated latitude bands. 
The ionization rate does not vary with time. 
It is axially symmetric around the solar rotation axis, and the latitudinal modulation is described using four parameters: the baseline value $\beta_0$, the amplification factor $f$, the heliographic latitude $\Theta$ of the amplification, and the latitudinal width of the amplified portion $\Delta \Theta$.
Setting $f = 1$ results in a spherically symmetric ionization.
Different combinations of these parameters will help us determine the influence of a localized increase in the total ionization rate on the light curve, depending on its strength and location. 

We start from looking at the variation in individual bands, then move to investigating the disturbances in two separate bands, and finally check if the light curves obtained for two individual bands can be superposed to reproduce a light curve simulated for two disturbed bands together.

The objective of this portion of our study is to verify the approach where the ionization rates are varied independently in heliolatitude bands covering the entire range of latitudes between the poles. 

\subsubsection{Variation of the ionization rate in individual latitude bands}
\label{sec:latiVaria1}
\noindent
In this experiment, we set $\beta_0=6 \times 10^{-7}$ s$^{-1}$, corresponding to the average observed value near the ecliptic, and vary the two parameters: $f$ and $\Theta$.
\citet{quemerais_etal:06b} found that during solar minimum, the total ionization rate based on SOHO/SWAN observations is of the order of $6-7 \times 10^{-7}$ s$^{-1}$, while during the solar maximum it can be twice as large.
We consider a similar range of values of the total ionization rate in this study.

\begin{figure*}[ht!]
\centering
\includegraphics[width=1.1\linewidth]{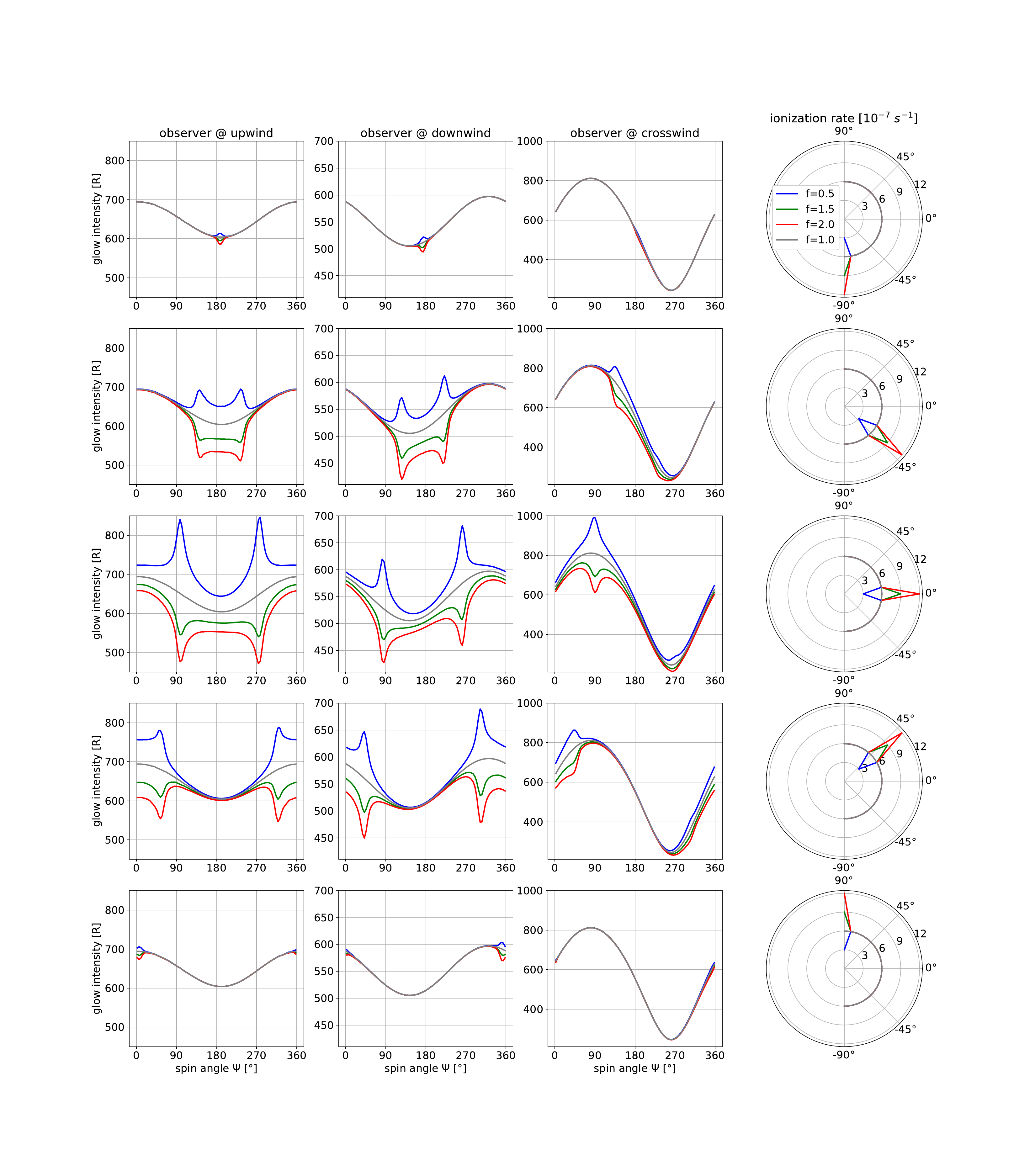}
\caption{Synthetic GLOWS light curves seen from 3 observer locations: upwind, downwind, and crosswind, for the ionization rates models presented in the extreme right column of panels.}
\label{fig:ion_stat_lat}
\end{figure*}

The latitudinal profile of the total ionization rate has been divided into bins of $\Delta \Theta =10\degree$ width.
We considered variations in the total ionization rate of a single bin at different heliographic latitudes (see the right column of the plots in Figure \ref{fig:ion_stat_lat}).
We present the results for the ionization rate disturbances at the solar equator ($\Theta=0 \degree$), both poles ($\Theta= \pm 90 \degree$), and two intermediate latitudes ($\Theta= \pm 40 \degree$).

For each $\Theta$, we simulated a reduction the ionization rate by half ($f=0.5$), an increase by a factor of 1.5 ($f=1.5$), and a doubling the total ionization rate at $\Theta$ ($f=2$).
The latitudinal profiles of all the studied combinations of the $\Theta$ and $f$ parameters are shown in the extreme right column of Figure \ref{fig:ion_stat_lat} as polar plots.
The ionization rates models have axial symmetry with respect to the axis of solar rotation.
The simulations were performed for the same upwind, downwind, and crosswind locations as in the spherically symmetric case, discussed in Section \ref{sec:ion_stat_lat}.

The first, second, and third column in Figure \ref{fig:ion_stat_lat} show the simulated GLOWS light curves seen by the observer placed at the upwind, downwind, and crosswind directions, respectively.
These results suggest that a change in the ionization rate at the solar poles has a minimal impact on the observed light curves (see the top and bottom rows).
This is primarily due to the small solid angle of the space affected by the altered ionization rate.
We consider this an important finding: the light curves are almost insensitive to variations of the ionization rate at the poles.

In the case of a disturbance in the ionization rate located at intermediate heliographic latitudes, the effect is much more significant (the second and fourth rows).
The magnitude of the change in the intensity scales approximately linearly with the magnitude of the ionization rate disturbance.
The spin angle range where the light curve is affected is much larger than the span of the affected heliolatitudes.
For a disturbance at mid-latitudes, approximately half of the light curve is affected.

The largest effect is observed when a disturbance of the ionization rate is located at the solar equator (see the third row in Figure \ref{fig:ion_stat_lat}).
In this case, practically the entire light curve reacts to these changes, albeit to a different extent for different spin angles.
Some asymmetries are observed, particularly noticeable from the crosswind position.
These asymmetries arise from the asymmetric distribution of the gas, which was also apparent in the previously discussed fully spherically symmetric case.

\subsubsection{Superposition of two latitudes.}
\label{sec:latiSuperposition}
\begin{figure*}[ht!]
\centering
\includegraphics[width=1.1\linewidth]{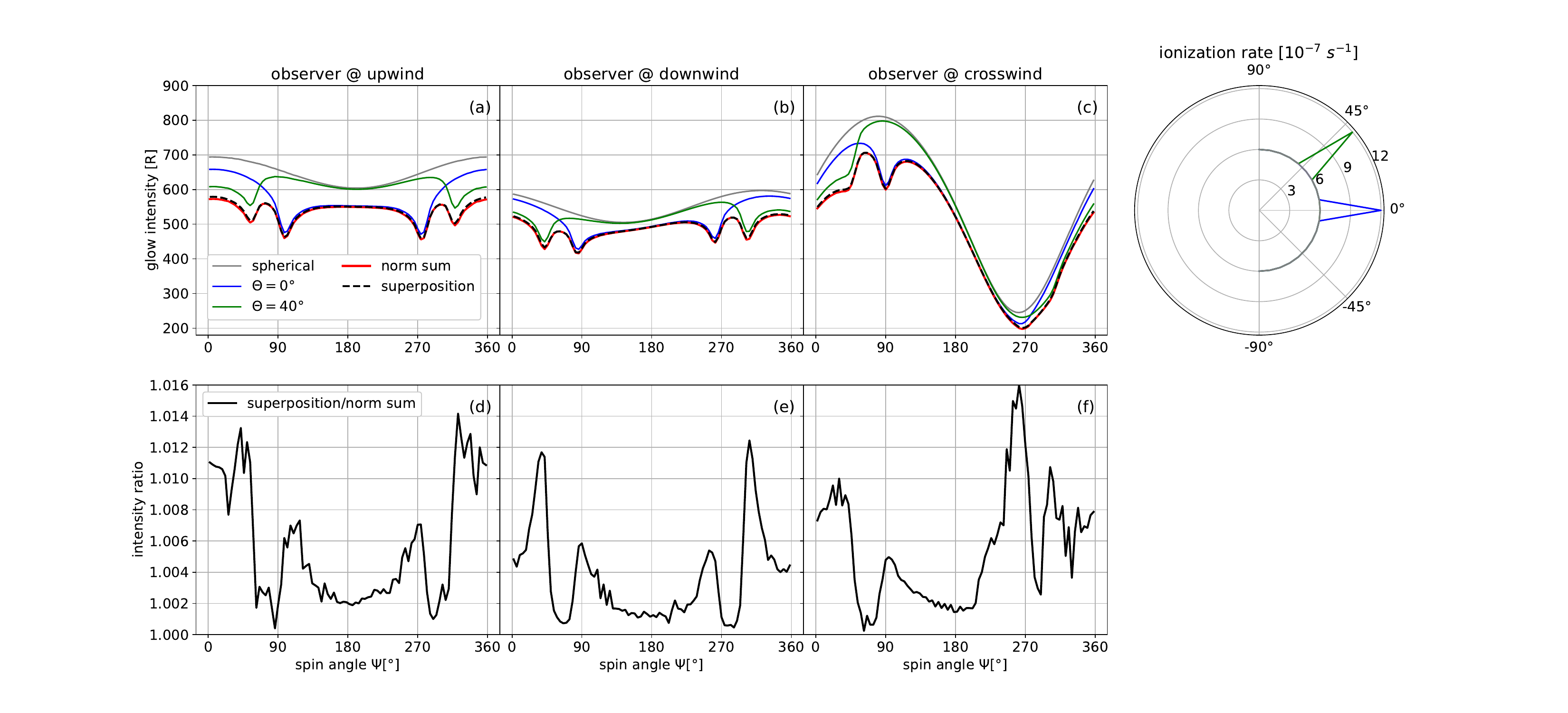}
\caption{Synthetic GLOWS light curves seen from 3 observer locations: upwind, downwind, and crosswind. Panels (a)---(c) show the simulated GLOWS light curves for different synthetic ionization rate  models: spherically symmetric (gray), a twofold increase of the total ionization rate at the equator (blue), a twofold increase of the total ionization rate at latitude $\Theta=40 \degree$ (green), a twofold increase of the total ionization rate at both the equator and latitude $\Theta=40 \degree$ (black dashed). The red line shows the normalized sum of the blue and green light curves.
Panels (d)---(f) show the ratio of the black dashed and red lines.
The polar plot on the right shows the ionization rate profiles for the simulated light curves, color-coded identically as the light curves.}
\label{fig:ion_lat_superposition}
\end{figure*}

\noindent
The last numerical experiment in this section was carried out for a superposition of two regions of increased ionization rate, located at heliographic latitudes $\Theta=0 \degree$ and $\Theta=40 \degree$.
The baseline total ionization rate was identical as previously, $\beta_0=6\times 10^{-7}$ s$^{-1}$, and the amplification factor was $f=2$.
The ionization rate model and the resulting light curves are presented in Figure \ref{fig:ion_lat_superposition}.
We denote the light curves simulated for this case as $I_{AB}$.
The objective was to verify if the light curves simulated for this model of ionization are similar to a composite light curve built of the light curves simulated for the corresponding disturbances at individual latitudes, as those discussed in Section \ref{sec:latiVaria1}.

For the two selected cases, A and B, where the ionization rate disturbances occur in individual latitudinal bands, we label the corresponding light curves as  $I_A$ and $I_B$. The case for the same baseline ionization rate and no disturbances is denoted as $I_0$.

Now, we can add these two light curves with individual ionization rate variations at latitudes of $\Theta=0 \degree$ and $\Theta=40 \degree$ (blue and green lines in Figure \ref{fig:ion_lat_superposition}) and subtract the spherical case (gray line in Figure \ref{fig:ion_lat_superposition}): 
$I_{A+B} = (I_A - I_0)+ (I_B - I_0) + I_0 = I_A + I_B - I_0$.
With this, we can compare a normalized light curve $I_{A+B}$ with the light curve $I_{AB}$ calculated from the model that includes ionization rate changes at both of these latitudes simultaneously.
Panels (a)---(c) illustrate this comparison for different observer positions.
Panels (d)---(f) present the ratio of the normalized sum of light curves with individual variations to the light curve calculated from the model containing ionization rate changes at both latitudes simultaneously $I_{A+B}/I_{AB}$.
Deviations of this ratio from 1 are below 2\% (see the lower row of panels in Figure \ref{fig:ion_lat_superposition}), indicating a very good agreement between both curves.
This suggests that with appropriate normalization, it is reasonable to construct simulated light curves for complex ionization rate profiles from ``building blocks'' of simulated light curves for ionization rate varied for individual latitudinal bins.

\subsection{Transient global decrease in the total ionization rate.}
\label{sec:ion_sw_time}
\noindent
So far, we have dealt with stationary cases, where variations in the ionization rate were constant over time.
Now, we explore effects of transient changes in the ionization rate.
We aim to investigate how quickly the helioglow reacts and how long the effects remain visible after the ionization rate returns to its original level.

\begin{figure*}[ht!]
\centering
\includegraphics[width=0.7\linewidth]{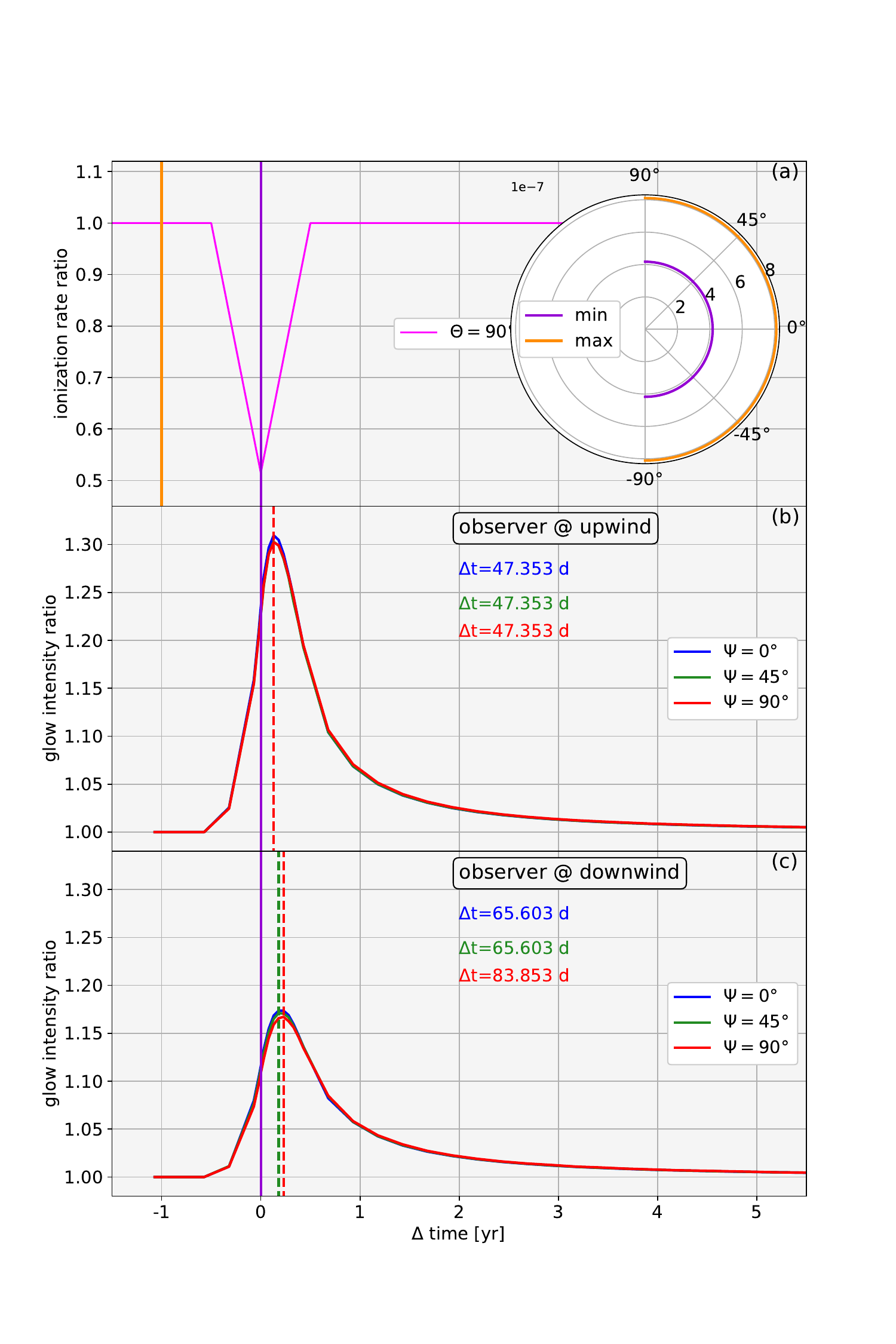}
\caption{Simulation results for spherically symmetric transient decrease of the total ionization rate. 
Panel (a): adopted time evolution of the total ionization rate. In the polar inset, the latitudinal profile is shown for two selected times, marked by the solid vertical lines. 
Panel (b) and (c): the ratios of the glow intensity for selected spin angles to the reference intensity before the considered ionization rate change, as seen by an observer in the upwind and downwind vantage points. 
Blue, green, and red colors mark the ratios of intensities at the spin angles indicated in the legends. 
The purple vertical bar marks the time of the largest reduction of the ionization rate.
The orange vertical bar corresponds to the reference epoch before the ionization rate disturbance.
Vertical dashed lines indicate the time when the selected bins of the light curve responded most strongly to the considered decrease in the total ionization rate.
The delays of the maxima of the intensity relative to the minimum of the ionization rate are listed as the $\Delta$ values for the spin angles shown.
}
\label{fig:ion_sw2D_time}
\end{figure*}

\begin{figure*}[ht!]
\centering
\includegraphics[width=0.7\linewidth]{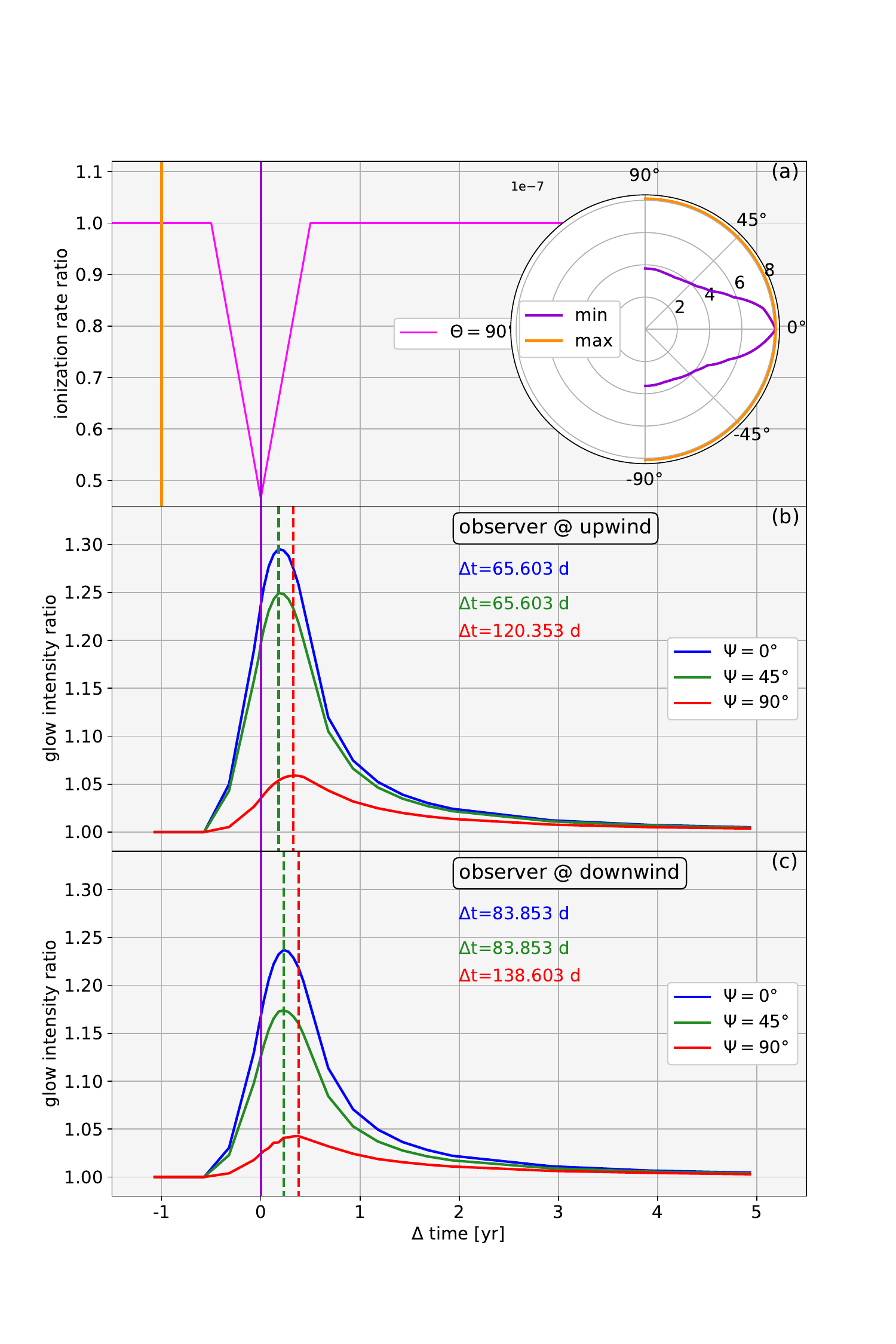}
\caption{Similar simulation results as in Figure \ref{fig:ion_sw2D_time}, but for an additional latitudinal modulation of the transient decrease of the total ionization rate.
}
\label{fig:ion_sw_time}
\end{figure*}

We have a certain ionization rate profile that at some moment begins to linearly decrease during six months and then returns to the initial state with the same change rate.
Here, we consider two situations: 
\begin{itemize}
\item The transient decrease is spherically symmetric (meaning that the total ionization rate decreases uniformly at the equator, pole, and every intermediate latitude), as shown in the first row in Figure \ref{fig:ion_sw2D_time}.
\item The transient decrease has different values for different heliographic latitudes (the latitudinal profile during the temporary increase corresponds to models of the solar wind structure during solar minimum \citep{porowski_etal:22a}, thus, we are dealing with a restructuring of the solar wind similar to that occurring during the solar-maximum-to-minimum transition, but on the yearly time scale). This case is shown in the first row in Figure \ref{fig:ion_sw_time}.
\end{itemize}

In both cases presented in Figures \ref{fig:ion_sw2D_time} and \ref{fig:ion_sw_time}, we observe a linear decrease in the ionization rate from the baseline level ($8.1 \times 10^{-7}$ s$^{-1}$) to a certain value over a period of six months, followed by a linear increase back to the initial level (see panel (a) in Figure \ref{fig:ion_sw2D_time} and \ref{fig:ion_sw_time}).
The response of the GLOWS light curves to the ionization change is shown in panels (b) and (c) for an observer placed near the upwind and downwind positions, respectively. 
We show the ratio of the helioglow intensity to the base value from the time before the ionization rate has changed.

The ISN H gas, which is the source of the helioglow, continuously flows into the vicinity of the Sun from the upwind direction. 
The flow speed is approximately 4--5 au per year.
After a change in the ionization rate, it takes some time for the gas to react.
In the case of a decrease in the ionization rate, this reaction results in a gradual increase of the gas density in the proximity of the Sun, and consequently, an increase in the intensity of the helioglow.
For an increase in the ionization rate, there is a gradual decrease in the gas density, which results in a decrease of the brightness of the helioglow. 
The time required for the reaction varies in different directions in the sky due to the non-uniform distribution of the gas in the heliosphere.

As a result, in all cases there is a time delay between the ionization rate drop (marked with a purple vertical line) and the maximum of the glow intensity (marked with dashed vertical lines of different colors for the selected spin angles) visible in our simulations.

In the case of the spherically symmetric decrease (see Figure \ref{fig:ion_sw2D_time}), the delay is just under 50 days for an observer located in the upwind direction and over 65 days for an observer in the downwind direction.
It is worth noting that for the downwind case, the delay for the light curve portion closest to the ecliptic plane ($\Psi=90 \degree$) is noticeably larger and reaches 83 days, even though its magnitude is the smallest of the three cases shown.

The case presented in Figure \ref{fig:ion_sw_time} exhibits a latitudinal structure that appears for a duration of one year.
The total ionization rate decreases to different values depending on the heliographic latitude.
Qualitatively, the profile shown in the polar-plot inset (panel (a), purple line) corresponds to the solar wind structure expected during the minimum of solar activity, while the baseline case, drawn with orange line, to that corresponding to the maximum of solar activity.

The latitudinally-structured variation in the ionization rate results in different time delays of the helioglow variation in different portions of the light curve, both for the upwind and downwind observers. The delay in the portion close to the ecliptic plane reaches 120 days or 138 days, depending on the vantage point.

The magnitude of the helioglow variation for a latitudinally-structured variation of the ionization rate strongly varies with the spin angle. 
The largest variation is observed in the ecliptic plane, and the smallest for spin angle corresponding to high latitudes. 
This behavior is consistent with the observation by \citet{bertaux_etal:96a} that an ecliptic darkening in the helioglow appears during the minimum of solar activity in the helioglow sky maps obtained from SOHO/SWAN observations, which disappears during times of high solar activity.

The intensity of the glow increases with a certain delay discussed above, but eventually returns to the level from before the change in ionization rate.
The time it takes for the glow to return to its initial state (with an accuracy of 1\%) is referred to as the relaxation time.
In the spherically symmetric case (Figure \ref{fig:ion_sw2D_time}), the relaxation time is around 3.5 years, regardless of the observer's position or spin angle.
In the latitudinally-structured case (Figure \ref{fig:ion_sw_time}), the relaxation time is about 2.5 years for spin angles around the ecliptic plane ($\Psi=90 \degree$) and just under 4 years for spin angles corresponding to higher ecliptic latitudes ($\Psi=0 \degree$ and $\Psi=45 \degree$).
Again, the position of the observer does not affect the estimated relaxation time.

Summing up this section, we found that a time variation in the ionization rate with an amplitude of f $\sim 2$ results in a brightening of the helioglow by factor up to $\sim 1.3$ for the upwind vantage point and $\sim 1.15$ for the downwind vantage point. 
The amount of brightening is similar for the entire light curve for a given vantage point.
The maximum of the brightening is delayed by 1.5 -- 3 months with respect to the maximum of the ionization rate. 
For a latitudinally structured variation of the ionization rate, the magnitude of the brightening is a strong function of the spin angle, and the delays in the maximum of the brightening are longer.
The relaxation time of the helioglow after a disturbance in the ionization rate is approximately twice longer than the rise time of the disturbance (2 years as compared with 1 year in the simulated case.)

\section{Transient increase of the Lyman-$\alpha$ radiation} 
\label{sec:lya_time}
\noindent
In this section, we investigate the influence of variations of the illumination of ISN H by the solar \lya{} emission on the GLOWS-like light curves. 
Similarly as we did in Section \ref{sec:ion_sw_time} for the ionization rate, we start with a spherically symmetric time variation of the solar \lya output in Section \ref{sec:lyaFlash}. Subsequently, in  Section \ref{sec:lyaFlashLati}, we analyze the effects of a small polar dimming of the solar \lya flux, pointed out by \citet{auchere:05}, \citet{pryor_etal:96, pryor_etal:01a}, and most recently by \citet{strumik_etal:21b, strumik_etal:24a}.

\subsection{Spherically symmetric flash of the solar \lya output}
\label{sec:lyaFlash}
\noindent
The solar \lya{} radiation affects the glow through two mechanisms.
The first one is a simple illumination effect, where the more photons illuminate hydrogen, the brighter the glow becomes.
The second one consists of a variation in the radiation pressure. 
With an increased flux of solar \lya photons, the resonant radiation pressure force acting on H atoms increases proportionally to the increase in the illuminating flux. 
The increased radiation pressure repels ISN H atoms stronger, thus reducing the density of ISN H near the Sun, which results in less gas available for excitation, and in effect in a decrease of the helioglow intensity. 
Here, we perform a simulation study to determine the balance between these two effects.

We performed simulations similar to those for the case of spherically symmetric time-variation of the ionization rate. 
We adopted a model of the spectral shape of the solar \lya line following \citet{IKL:18a, IKL:20a}.
With this, the sensitivity of radiation pressure on radial velocity of individual atoms is acknowledged.

\begin{figure}[!ht]
\centering
\includegraphics[width=0.7\linewidth]{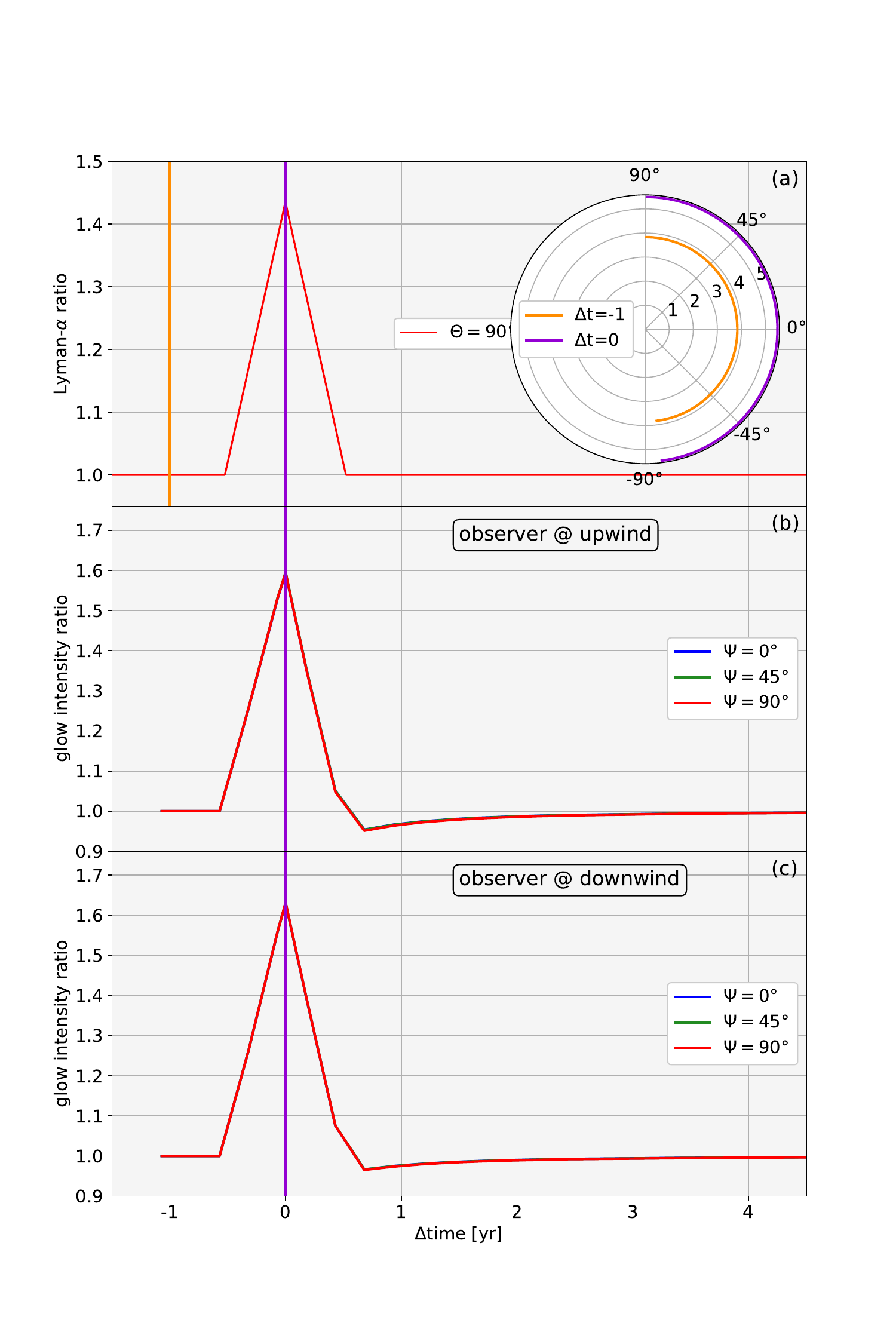}
\caption{Simulation results for a transient increase of the \lya{} radiation. Panel (a): time evolution of the \lya{} irradiance. In the polar inset, the latitudinal profile is shown for two selected times, marked by the solid vertical lines. The solid lines on polar plot show latitudinal profiles of the total solar irradiance [$10^{11}$ ph s$^{-1}$ cm$^{2}$].
Panels (b) and (c): ratios of the glow intensity in selected bins of the light curve to the reference intensity before the considered radiation change as seen from the upwind and downwind directions.}
\label{fig:lya_time_2D}
\end{figure}

In the model of transient brightening of the Sun, we assumed that the total solar \lya flux increases during 6 months from a magnitude used in the simulations discussed in Section \ref{sec:ion_models}, i.e., $3.8 \times 10^{11}$ cm$^{-2}$ s$^{-1}$, corresponding to values typical for solar minimum, to $5.5 \times 10^{11}$ cm$^{-2}$ s$^{-1}$, which is characteristic for solar maximum. After reaching the peak value, the magnitude of the solar flux drops to the original level, as shown in panel (a) of Figure \ref{fig:lya_time_2D}. The ionization rate is adopted as spherically symmetric, identical with the pre-variation magnitude shown in Figure \ref{fig:ion_sw2D_time}.

It must be noted that variation of the total flux in the solar \lya line is not uniform for all wavelength within the line. 
For radiation pressure and the helioglow, the most relevant portion in general is that within approximately $\pm 40$ \kms due to the Doppler effect (which corresponds to $\Delta \lambda \pm 0.016$ nm). 
The exact value cannot be determined because it depends on the location along the line of sight, the geometry of the line of sight, etc. 
According to the model of the evolution of the solar \lya profile by \citet{IKL:20a}, the amplitude of variations of the total flux during the solar cycle is amplified in the central portion of the profile.
This variation affects both illumination, and thus the total intensity of the helioglow, and radiation pressure.
In the adopted model, for the ratio of irradiance $5.5 \times 10^{11}/3.8 \times 10^{11}=1.45$, the radio of spectral fluxes at the line center is equal to 1.69, and the ratio of spectral fluxes averaged over $\pm 40$ \kms to 1.62. This latter ratio is very close to the magnitude of helioglow brightening at the peak in Figure \ref{fig:lya_time_2D}.

Panels (b) and (c) of Figure \ref{fig:lya_time_2D} show the time variation of the helioglow intensity in the selected portions of the light curves for the \lya impulse.
The main part of the response is immediate, as we observe an increase in the glow intensity when the illumination increases.
This is because the propagation of the solar flux and excitation of H atoms are immediate in the time scale of the problem. 
However, the variation of the intensity is not symmetric in time, unlike the \lya impulse itself.
After the impulse is over, the helioglow intensity drops a few percent below the magnitude characteristic for the pre-impulse times. 
This is because the increasing part of the time evolution of the light curve is affected by the increase in the illumination, while the density distribution of ISN H is not yet affected. The effect of the increased radiation pressure takes more time to accumulate and thus becomes visible later in the time evolution. 
Later on, the density remains depleted after some time after the impulse is over, which results in the small drop of the helioglow intensity that persists approximately a year.
This behavior is visible regardless of the vantage point of the observer (upwind or downwind).

A conclusion from this part is that in the case of solar illumination variation, the reaction of the helioglow is instantaneous, but the helioglow brightening is a little reduced relative to the magnitude of the illumination increase due to the mitigation effect of radiation pressure, which counteract the brightening. 
This latter effect is also responsible for a small reduction in the helioglow intensity after the illumination brightening is over.
Relaxation of this effect takes approximately a year. 
Moreover, the increase in helioglow intensity is larger than the increase in the \lya irradiance because the central part of the \lya line is more variable than the whole total solar irradiance.

\subsection{Variation of a latitudinally-anisotropic \lya emission}
\label{sec:lyaFlashLati}
\begin{figure}[!ht]
\centering
\includegraphics[width=0.7\linewidth]{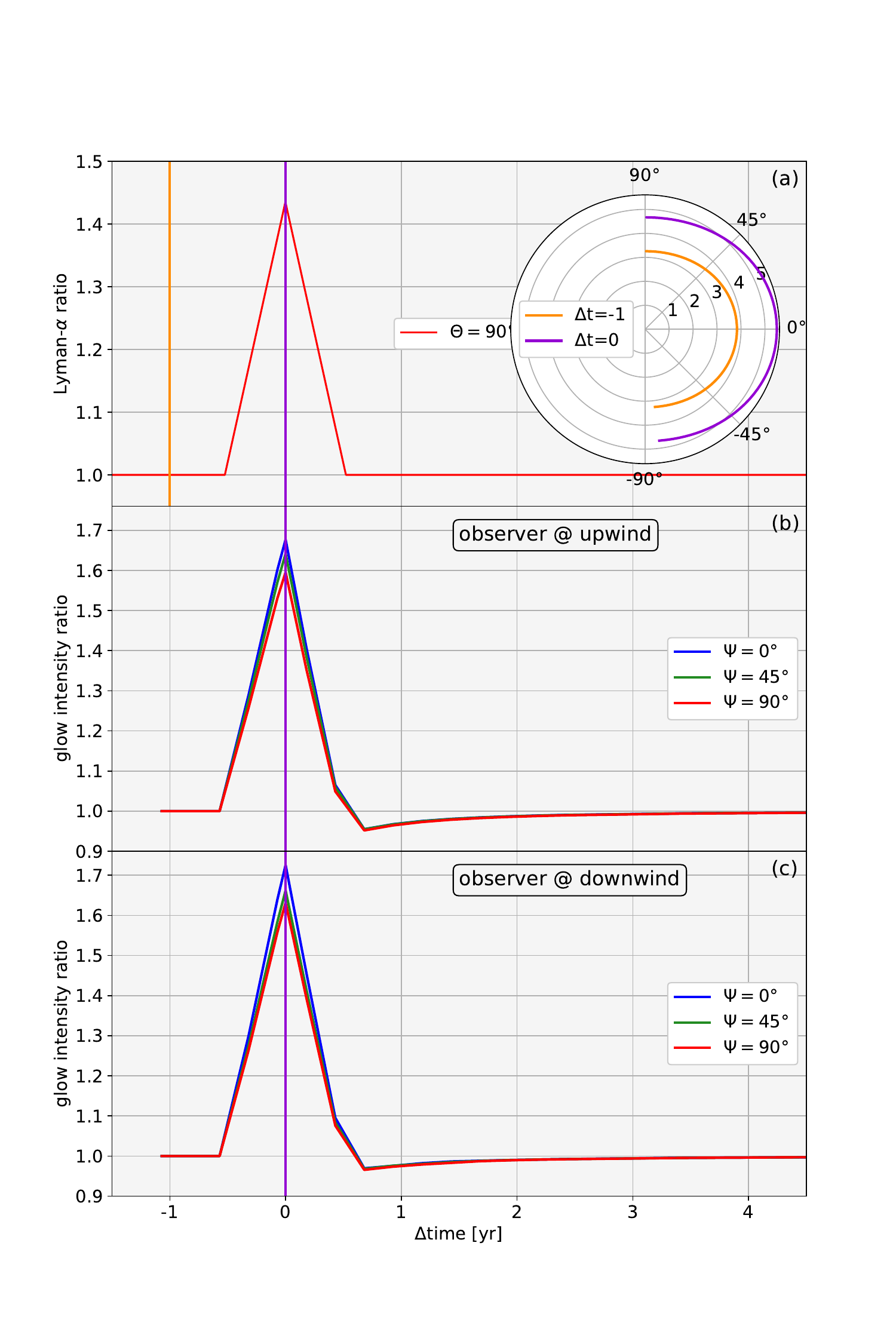}
\caption{Similar simulation results as in Figure \ref{fig:lya_time_2D}, but for an additional latitudinal modulation of the transient decrease of the total \lya{} radiation caused by the polar-equator anisotropy.}
\label{fig:lya_time_3D}
\end{figure}

\noindent
In this part of the research, we add a latitudinal anisotropy of the solar \lya flux. 
We assume that the total flux varies with heliolatitude according to Equation 29 in \citet{kubiak_etal:21a}: $I_{\text{tot}}(\phi) = I_{\text{tot}}(0)(a \sin^2 \phi + \cos^2 \phi)$, where $I_{\text{tot}}$ is the solar \lya irradiance. 
As a result, the solar irradiance in the \lya line at the poles is equal to $a = 0.85$ of that at the equator. 
This is illustrated in the inset in panel (a) in Figure \ref{fig:lya_time_3D}.
The magnitude of this anisotropy is on the order of the anisotropy obtained by \citet{strumik_etal:24a} based on analysis of SWAN photometric maps of the helioglow, in particular for the epochs of high solar activity. 

With these modifications to the solar illumination, we repeat the simulations presented in Section \ref{sec:lyaFlash}.
The results are shown in panels (b) and (c) of Figure \ref{fig:lya_time_3D}.

\begin{figure}[!ht]
\centering
\includegraphics[width=0.5\linewidth]{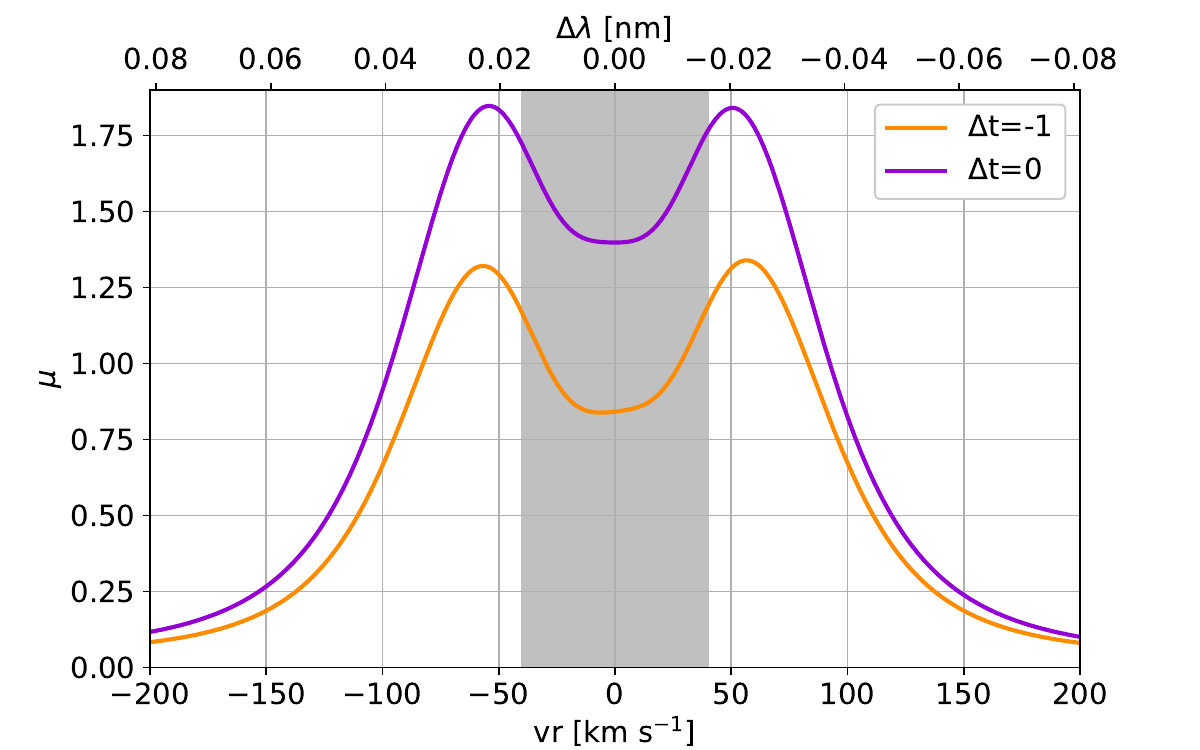}
\caption{\lya profiles before the increase of the illumination (orange solid line) and for the peak (violet solid line). The shadowed area between radial velocities $\pm 40$  km s$^{-1}$ shows the part of the profile that is responsible for the helioglow formation and radiation pressure effect. The composite solar irradiance delivered by LASP \citep{machol_etal:19a} is an integral over 1 nm bin between wavelengths 121 and 122 nm.}
\label{fig:lya_profile}
\end{figure}

In the case with polar-equator anisotropy, we observe an additional latitudinal effect, where different parts of the light curve are differently affected (see blue, green and red lines in panel (b) of Figure \ref{fig:lya_time_3D}.
The largest brightening relative to the corresponding intensity magnitude before the flash occurs for the polar phase angles ($\Psi=0 \degree$), the smallest for the ecliptic angles ($\Psi=90 \degree$). 
The magnitude of the brightening for the ecliptic angles is very similar to that obtained for the spherically-symmetric illumination and it's around 1.6, whereas total irradiance in \lya ratio is near 1.4.
This is because the helioglow is sensitive to the central part of the \lya profile ($\sim \pm 40$ km s$^{-1}$), shown as a grey shaded area in Figure \ref{fig:lya_profile}, which features a larger  modulation than that for the composite solar \lya irradiance.

The time evolution of the light curves, in particular the decay and relaxation times, is very similar to those obtained for the spherically-symmetric case.

The conclusion from this portion is that latitudinal anisotropy of the solar illuminating flux does not modify the timing of the evolution of light curves, but modifies the pole-to-equatorial ratios of the helioglow intensity. 
Since the portion of the solar \lya line responsible for the actual illumination of the gas and the radiation pressure is very narrow and concentrated close to the nominal \lya wavelength, the anisotropy of the actual illumination is greater than the anisotropy of the irradiance, i.e., the wavelength-integrated \lya flux.

\section{Discussion and conclusions}
\label{sec:summary}
\noindent
WawHelioGlow is an optically thin model.
We believe that for photometric observations performed from 1~au, this is a sufficient approximation based on 
recent comparisons between the WawHelioGlow model with SOHO/SWAN observations, which show a good agreement, even though multiple scattering have not been included \citep{strumik_etal:21b}.
That comparison was done for full WawHelioGlow model, which uses realistic models of ionization rate and radiation pressure dependent on time, latitude, and atom velocity.
While \citet{quemerais_bertaux:93a} and \citet{quemerais:00} suggested that multiple scattering effects must be taken into account also for observations performed at 1 au, they neglected the time- and heliolatitude variations of the ionization rate, illumination and radiation pressure, as well as the dependence of the illumination and radiation pressure on radial speeds of individual atoms. 
We also found that the maximum of the source function for the \lya signal from the hydrogen glow that we observe near Earth's orbit is within a few au from the Sun \citep[see Figures 4, 6--9 in][]{kubiak_etal:21b}.
This is inside the optically thin region \citep{IKL:22a}.
Therefore we tentatively adopt the optically thin model.
A need to extend WawHelioGlow into multiple scattering will be considered after more detailed comparison of the  available observations with model predictions are performed, which is a subject of ongoing work. 
This, however, is outside the scope of this paper.

The anisotropic distribution of the incoming ISN H, the complex structure of the solar wind, and the non-uniform illumination make the intensity of the helioglow emissions behave in a non-trivial manner.
The analysis we have done aims to help decompose the effects of the complex structure of the total ionization rate into simpler factors and investigate the impact of each of them on the future GLOWS light curve.
One of the goals was to asses which portions of the latitudinal ionization rate profile is the most important, how soon the helioglow reacts on changes in the ionization rate or solar \lya illumination, and how long the modifications of the light curves last in comparison with the duration of the ionization rate and illumination variations.
This will allow for a better understanding of realistic light curves observed by the GLOWS instrument.
We have examined four types of disturbances that affect the shape of the light curve:
\begin{enumerate}
\item Changes of a spherically symmetric ionization rate, constant in time. We found that the helioglow intensity is reduced with an increase of the global ionization rate approximately logarithmically, with the parameters of this logarithmic dependence being a function of the vantage point and spin angle.
\item Stationary cases where ionization rate featured an increase or decrease within a narrow latitudinal band at selected heliographic latitudes.
Within this case, we examined different positions, as well as different magnitudes of the change.
We found that changes of the ionization in the polar regions have a very small impact on the light curves, limited to spin angles near 0$\degree$ and 180$\degree$ depending on the pole (N or S).
The closer the change is located to the solar equatorial plane, the greater the angular range of the light curve affected by this effect.
In the case of the equatorial-plane changes in the total ionization rate, the entire light curve is substantially affected, although to a different extent in its different segments.
\item A global but transient decrease in the total ionization rate.
Here we considered both a spherically symmetric case and one with a specified latitudinal structure, where different heliographic latitudes experienced a decrease of varying amplitude.
Simulations showed that the light curve responds with some delay to the decrease in ionization rate.
This delay depends on both the observer's position and the heliographic latitude.
The fastest response of the light curve occurs after 47.4 days for the symmetric case when the observer is in the upwind position.
The longest delays are observed for the observer in the downwind position near the ecliptic plane in the case with latitudinal structure, reaching up to 138.6 days.
One should note that the considered range of variations of the total ionization rate from $\sim$4 to $\sim$8$\times10^{-7}$ s$^{-1}$ implies time scales from $\sim$14 to $\sim$28 days.
This gives a lower bound on the characteristic time scale for the processes of rebuilding the hydrogen distribution around the Sun due to the total ionization rate variations.
Our results agree with this estimate, showing larger but similar (in terms of the order of magnitude) delays in the response of the helioglow to the total ionization rate transients.
\item A global, transient change in the intensity of the solar \lya radiation, without a change in the total ionization rate (only the solar illumination and radiation pressure have changed).
Here we considered spherically symmetric case, as well as anisotropic one where there is more radiation seen from the equator than from the poles.
An increase in the radiation intensity results in an almost immediate increase in the intensity of the helioglow.
However, the following decrease in the intensity is not symmetric and reaches a lower level than the initial one.
Within the following few years, the intensity returns to its initial values.
\end{enumerate}

We also found an effect of relaxation in the helioglow after transient disturbances are over. 
The relaxation time for the hydrogen glow is shorter in the case of \lya irradiance change (almost 3 years), than it is in the case of total ionization rate change (almost 4 years), except for spin angles close to the ecliptic plane, where intensity of the glow returns to the initial levels after about 2.5 years.

We verified that the strategy of fitting the observation using a model of latitudinal structure of the ionization rate approximated by freely--varying magnitudes of the rate in individual latitudinal bands is viable in the sense that one can superpose the lightcurves simulated for different magnitudes of the ionization rate in individual bands. But we also found that the GLOWS-like lightcurves are very little sensitive to variations in the ionization rate in polar bands.
This makes this strategy unsuitable for determining the magnitude of the ionization rate at the solar poles. 
Instead, an alternative strategy needs to be employed, where the profiles of the ionization rate are approximated by global functions of heliolatitude with free parameters, and retrieving the polar rates is possible by evaluating these functions with the fitted parameters. 
This latter approach is based on the notion that the solar wind with its latitudinal structure is a global phenomenon and the solar wind parameters at all heliolatitudes are intrinsically related to each other, differently for different phases of the solar activity cycle.

In the reality, variations in the illumination and the ionization rate operate simultaneously, so their effects are intertwined. 
Our findings will be considered in the construction of algorithms of retrieval of the ionization rate profiles and heliolatitude dependence of solar wind parameters based on future observations of the helioglow by GLOWS.
Our findings can be applied for any observations of the helioglow similar to those planned for GLOWS, including an appropriate subset of publicly available observations from SOHO/SWAN \citep{bertaux_etal:95} that we are currently working on. 

\emph{Acknowledgments}\\
This research was facilitated by the International Space Science Institute (ISSI) in Bern, through ISSI International Team project \#541 (Distribution of Interstellar Neutral Hydrogen in the Sun's Neighborhood).
The authors would like to express their special thanks to Dimitra Koutroumpa for extremely helpful discussions within and outside this ISSI team.
This study was supported by Polish National Science Center grants 2019/35/B/ST9/01241, 2018/31/D/ST9/02852, and by Polish Ministry for Education and Science under contract MEiN/2021/2/DIR.



\bibliographystyle{aasjournal}
\bibliography{iplbib}


\end{document}